\documentclass[journal=jacsat,manuscript=article]{achemso}

\usepackage[T1]{fontenc} % Use modern font encodings
\usepackage{braket}
\usepackage{chemmacros} 
\usepackage[version=4]{mhchem}
\usepackage{multirow}
\newcommand{\kcal}[0]{kcal mol$^{-1}$}
\renewcommand{\cal}[0]{cal mol$^{-1}$}
\newcommand{\kJmol}[0]{kJ mol$^{-1}$}
\newcommand{\Qf}[0]{Q$_\text{f}$}
\newcommand{\pQL}[0]{(Q)$_\Lambda$}
\newcommand{\EH}[0]{E$_\text{H}$}

\author{James H. Thorpe}
\affiliation{Department of Chemistry, Southern Methodist University, Dallas, TX 75275, USA}
\author{Zachary W. Windom}
\affiliation{Quantum Theory Project, University of Florida, Gainesville, Florida 32611, USA}
\author{Rodney J. Bartlett}
\email{bartlett@qtp.ufl.edu}
\affiliation{Quantum Theory Project, University of Florida, Gainesville, Florida 32611, USA}
\author{Devin A. Matthews}
\email{damatthews@smu.edu}
\affiliation{Department of Chemistry, Southern Methodist University, Dallas, TX 75275, USA}

\title[]
  {Factorized Quadruples and a Predictor of Higher-Level Correlation in Thermochemistry}

\keywords{}

\begin{document}

\begin{abstract}
    Coupled cluster theory has had a momentous impact on the ab initio prediction of molecular properties, and remains a staple ingratiate in high-accuracy thermochemical model chemistries. However, these methods require inclusion of at least some connected quadruple excitations, which generally scale at best as $\mathcal{O}(N^9)$ with the number of basis functions. It very difficult to predict, a priori, the effect correlation past CCSD(T) has on  a give reaction energies. The purpose of this work is to examine cost-effective quadruple corrections based on the factorization theorem of many-body perturbation theory that may address these challenges. We show that the $\mathcal{O}(N^7)$, factorized CCSD(T\Qf{}) method introduces minimal error to predicted correlation and reaction energies as compared to the $\mathcal{O}(N^9)$ CCSD(TQ). Further, we examine the performance of Goodson's continued fraction method in the estimation of CCSDT\pQL{} contributions to reaction energies,  as well as a ``new'' method related to \%TAE[(T)] that we refer to as a scaled perturbation estimator. We find that the scaled perturbation estimator based upon CCSD(T\Qf{})/cc-pVDZ is capable of predicting CCSDT\pQL{}/cc-pVDZ contributions to reaction energies with an average error of 0.07 \kcal{} and a RMST of 0.52 \kcal{} when applied to a test-suite of nearly 3000 reactions. This offers a means by which to reliably ballpark how important post-CCSD(T) contributions are to reaction energies while incurring no more than CCSD(T) formal cost and a little mental math.
\end{abstract}

\begin{tocentry}
\includegraphics{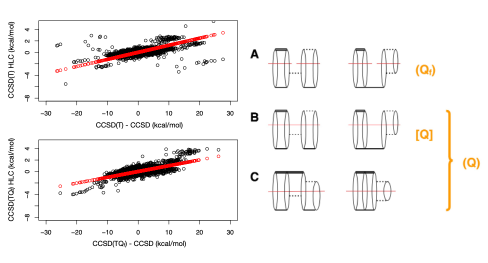}
\end{tocentry}

\section{Introduction}
Coupled cluster (CC) theory provides a systematically improvable route toward capturing the instantaneous electron correlation of small to medium-sized molecules. By modulating the maximal rank of the cluster operator according to calculation affordability and/or the desired accuracy, a hierarchy of CC methods are available albeit using algorithms that scale as increasingly high-order polynomials. This feature of CC theory has made it the method of choice for so-called composite model chemistries, which describe procedures by which predict experimental measurements of molecular properties such as bond energies or vibrational frequencies to a given degree of accuracy/precision/trueness via computed, or easily obtained, values.\cite{Karton-ARCC-2022} The exact details of which coupled-cluster contributions are included with which basis sets depends on the goals of the model chemistry in question. High-accuracy model chemistries, which seek to determine molecular enthalpies of formation to within 1 \kJmol{}, such as HEAT\cite{Tajti-JCP-2004, Bomble-JCP-2006, Harding-JCP-2008, Thorpe-JCP-2019}, W$n$\cite{Martin-JCP-1999, Boese-JCP-2004, Karton-JCP-2006, Karton-JCP-2012, Sylvetsky-JCP-2016, Semidalas-JPCA-2024}, FPA\cite{Csaszar-JCP-1998, East-JCP-1993, Kenny-JCP-2003, Schuurman-JCP-2004, Jaeger-JCTC-2010}, FPD\cite{Ruscic-JPCA-2000, Ruscic-JPCA-2002, Feller-JCP-2008, Peterson-TCA-2012, Feller-MP-2012, Dixon-ARCC-2012, Feller-TCA-2014}, and others (see Ref.~\citenum{Karton-ARCC-2022}), generally include not just CCSD(T)\cite{Purvis-JCP-1982,Raghavachari-CPL-1989} with a sizeable basis set, but also some amount of instantaneous four-electron correlation via CCSDT(Q)\cite{kucharski1989coupled,kucharski1993coupled,Bomble-JCP-2005}, CCSDTQ\cite{kucharski1992coupled}, or, more recently, CCSDT\pQL{},\cite{kucharski1998sixth,Bomble-JCP-2005} typically in a cc-pVDZ basis set. On the other hand, model chemistries that seek chemical accuracy (one \kcal{} within experimental measurements), such as G-$n$\cite{Pople-G1-1989, Curtiss-JCP-1990, Curtiss-JCP-1991, Curtiss-JCP-1998, Curtiss-JCP-2007} and CBS\cite{Petersson-JCP-1988, Petersson-JCP-1991a, Petersson-JCP-1991b, Montgomery-JCP-1994, Ochterski-JCP-1996, Montgomery-JCP-1999}, among others\cite{Karton-ARCC-2022}, generally stop at CCSD(T) with a relatively small basis set, as they target systems for which CCSD(T) itself is already an expensive calculation. While it is recognized that the neglect of post-(T) corrections, often referred to as higher-level correlation (HLC) corrections, in multireference systems is dangerous, it is also true that the absence of these terms in a model chemistry aiming for chemical accuracy introduces errors that, even for well behaved species, may put the goal of sub-\kcal{} accuracy in danger when combined with other approximations that must be made to reduce cost. Further, it is difficult to tell, quantitatively, when this will be the case a priori. The purpose of this work is to investigate some routes towards estimating the size of this post-CCSD(T), non-relativistic electron correlation, without incurring any increase in formal cost beyond $\mathcal{O}(N^{7})$ already expended in CCSD(T). 

This effort naturally begins with perturbation theory. Initial attempts to increase the accuracy of CCSD\cite{Purvis-JCP-1982} centered on accounting for the effect of the ${T}_3$ triple electron excitation operator, either perturbatively\cite{Purvis-JCP-1982,Raghavachari-CPL-1989,kucharski1998noniterative,crawford1998investigation,Stanton-CPL-1997,taube2008improving} or iteratively\cite{Urban-JCP-1985,noga1987towards,lee1984coupled}. These efforts eventually culminated in the $\mathcal{O}(N^8)$ CCSDT method.\cite{noga1987full} Pursuit of increasingly accurate calculations lead to similar attempts at incorporating quadruple electron excitations into the underlying CC framework.\cite{noga1989coupled,kucharski1989coupled,kucharski1993coupled,Bomble-JCP-2005,eriksen2014lagrangian} The groundwork for perturbative treatment of quadruple excitation operators like [Q], (Q), and (Q)$_{\Lambda}$ were originally postulated by Kucharski and Bartlett\cite{kucharski1993coupled,kucharski1998sixth}, which naturally lead to the $\mathcal{O}(N^{10})$ CCSDTQ method.\cite{kucharski1991recursive,kucharski1992coupled}. A later derivation by Bomble and Stanton used L\"{o}wdin's partitioning on a CCSDT reference to derive the versions of CCSDT(Q) and CCSDT\pQL{} that have been employed in high-accuracy thermochemistry, and resolved ordering issues similar to those that plague the definitions and performance of +T, [T], and (T).\cite{Bomble-JCP-2005, Bomble-JCP-2006} Of particular note is that perturbative triple corrections like [T] or (T), and perturbative quadruple corrections like [Q] or (Q), are $\mathcal{O}(N)$ cheaper than their counterparts that are complete through a given cluster operator rank. An important consequence of this reduced scaling is the capability not only to address larger molecular systems, but also to employ larger atomic basis sets in higher-level correlation calculations of small systems when exceedingly high accuracy is called for.\cite{Thorpe-JCP-2021, Thorpe-PCCP-2023}

However, in spite of the order $\mathcal{O}(N)$ in savings afforded by using methods like CCSDT\pQL{} (which scales as $\mathcal{O}(N^9)$) as compared to full CCSDTQ, this is a far cry from the $\mathcal{O}(N^7)$ cost of CCSD(T). The first step to reduce this cost is to remove the need to perform the iterative CCSDT component of these methods (an $\mathcal{O}(N^8)$ procedure), typically by replacing the converged $T_3$ CC amplitudes with a denominator-weighted contraction of $T_2$ with the Hamiltonian, as is done in CCSD(T). The remaining cost then arises from contractions with the Fock denominator, $D_4$---in this case, a 8-index tensor---used to construct an approximation to the $T_4$ operator. These methods are typically refereed to as CCSD(TQ).\cite{kucharski1998sixth} A reasonable way to reduce the algorithmic scaling, and the central point of this manuscript, focuses on eliminating the 8-index, $T_4$ denominator by invoking the factorization theorem of MBPT.\cite{frantz1960many,bartlett1975some} This allows for the elimination of the $D_4$ denominator in exchange for a product of two, 4-index denominators (e.g. $D_2^AD_2^B$), which is the idea used to define the \Qf{} correction\cite{Kucharski-AQC-1986,kucharski1998efficient,kucharski2010connected} that facilitates an estimate of connected quadruples excitations at $\mathcal{O}(N^7)$ cost. Initial studies of these factorized methods focused on a small selection of molecules, with the general conclusion being that negligible error was introduced into the calculation by adopting the \Qf{} method over analogous, but more expensive quadruple corrections.\cite{kucharski1998noniterative,kucharski1998sixth,kucharski1999connected,kucharski2001toward,musial2005critical,musial2010improving} Later benchmarks studied the performance of a variety methods in the prediction of ground state correlation energies.\cite{Eriksen-JCP-2016a, Eriksen-JCP-2016b} However, a robust analysis of these factorized approaches centered in the realm of accurate thermochemistry remains largely unexplored. 

After establishing the accuracy and characteristics of this factorization approximation, this work will investigate how a low-order approximation of the importance of the ${T}_4$ operator might be correlated with the size of post-CCSD(T) contributions of the more complete CCSDT(Q)$_\Lambda$ method\cite{kucharski1993coupled,kucharski1998sixth,Bomble-JCP-2005}, which has recently been demonstrated as a very accurate higher-level correlation approximation.\cite{Martin-MP-2014, Thorpe-Dissertation-2022, Thorpe-PCCP-2023, Franke-JPCA-2023, Semidalas-JPCA-2024, Spiegel-MP-2024} Although several indices have been proposed to indicate a strong deficiency in the zeroth-order approximation to the wavefunction\cite{Karton-JCP-2006, Martin-MP-2014, lee1989diagnostic, lee1989theoretical, ivanov2005new, bartlett2020index, faulstich2023s}, which usually coincides with large HLC corrections, there are limited means by which to correlate these metrics with a quantitative estimate of post-CCSD(T) contributions in thermochemistry. Approaches based on Goodson's continued fraction formalism have seen some success in the prediction of post-CCSDTQ correlation,\cite{Goodson-JCP-2002, Karton-JCP-2007, Feller-JCP-2007} along with estimates bested upon the percentage of perturbative triples correlation in the total atomization energy of a molecule, the well known \%TAE[(T)] metric.\cite{Karton-JCP-2006}

The remainder of this manuscript is organized as follows. First, we briefly introduce the theory behind the CCSD(T)-scaling methods that are the focal point of this work. We then describe how these methods are benchmarked against a subset of the W4.17 dataset,\cite{Karton-JCC-2017} and how this dataset is expanded to included several thousand chemical reactions. We then use these test-suites to analyze the errors incurred by factorizing the quadruple excitation contributions of these CCSD(T)-scaling methods, along with the size of the correlation missing between these methods and CCSDT\pQL{}, in the context of both raw correlation energies and reaction energies. Finally, we present a novel predictor of post-CCSD(T) higher-level correlation based upon these factorized quadruples methods that serves as a reliable guide as to when HLC effects need to be accounted for in a model chemistry.

\section{Theory}

\subsection{Coupled Cluster Theory and Perturbative Quadruples Corrections}
Coupled cluster theory is an exponential paramterization of the wavefunction 
\begin{align}
    \ket{\Psi}=e^T\ket{0}
\end{align}
for some single reference Slater determinant $\ket{0}$. The cluster operator, $T$, is expressed in terms of a sum of individual operators responsible for single, double, $\cdots$, up to $n$-fold electron excitations
\begin{align}
    T=T_1+T_2+\cdots+T_n
\end{align}
In the limit where the cluster operator is left untruncated, the FCI wavefunction is rigorously recovered. Otherwise, a truncation point is chosen to include up to a maximal rank cluster operator, $T_k$ defined as
\begin{align}
    T_k=(k!)^{-2}\sum t_{ij\cdots}^{ab\cdots}a^{\dagger}b^{\dagger}\cdots ji
\end{align}
The definition for the normal-ordered Hamiltonian is
\begin{align}
    H_N &= \sum_{pq} f_{pq} \{p^{\dagger}q\}+ \frac{1}{4}\sum_{pqrs}\braket{pq||rs}\{ p^{\dagger}q^{\dagger}sr\} \\
        &= f_N + W_N
\end{align}
which is expressed in terms of one and two body integrals. In this work we assume a canonical (Hartree--Fock) reference determinant such that $f_{pq} = \delta_{pq}\epsilon_p$, with $\epsilon_p$ being the orbital energies.

As there is some ambiguity in the literature, we wish to define the pertinent quadruple corrections that are the focal point of this study in the interest of clarity. 
We adopt the nomenclature [Q] to mean the energy correction associated with quadruple excitations correct through fifth-order in MBPT,\cite{kucharski1998noniterative} given by 
\begin{align}\label{eq:sqrBrakQ}
    \Delta E_{[Q]}^{[5]}=\braket{0|T_2^{\dagger}W_NT_4^{[3]}|0}
\end{align}
where the approximation to $T_4$ is correct through third-order in MBPT and only includes connected (C) diagrams
\begin{align}\label{eq:T4eqn}
    T_4^{[3]}=R_4\left( W_N\frac{T_2^2}{2} + W_NT_3\right)_C
\end{align}
The resolvent operator, $R_n$, is defined as
\begin{align}
    R_n(X)=(n!)^{-2}\sum \frac{\braket{\Phi_{ij\cdots}^{ab\cdots}|X|0}}{\epsilon_i+\epsilon_j+\cdots - \epsilon_b - \epsilon_a} a^{\dagger}b^{\dagger}\cdots ji
\end{align}
and is used to enforce the projection of a pure excitation operator onto the proper subspace. For example, with the double excitation cluster operator $T_2$ represented as a graphical fragment with external indices $A = \{a,b,i,j\}$, then $R_2(T_2) = \frac{1}{4}\sum_{abij}\frac{t^{ab}_{ij}}{\epsilon_i+\epsilon_j - \epsilon_b - \epsilon_a} a^\dagger b^\dagger ji$, which is the same operator with amplitudes divided by denominators $D_2^A = \epsilon_i+\epsilon_j - \epsilon_b - \epsilon_a$. Informally, we can also write the action of the resolvent as, for example, $R_2(T_2) = D_2^{-1} T_2$, and similarly, with a modified definition, the resolvent can be applied to de-excitation operators such as $R_2(T_2)^\dagger = T_2^\dagger D_2^{-1}$.

At this point, it is prudent to briefly review the tenets of the factorization theorem in MBPT. We first note that the first-order approximation of $T_2$ is
\begin{align}
    T_2 \approx T_2^{(1)} = R_2(W_N) = D_2^{-1} W_N
\end{align}
This allows \eqref{eq:sqrBrakQ} to be rewritten through fourth order as 
\begin{align}
    \braket{0|W_N D_2^{-1} W_N T_4^{[3]}|0}
\end{align}
The two $W_N$ operators must be disconnected since they only contain double de-excitation components. Thus, the external indices spanned by the quadruple excitation fragment $T_4^{(3)}$ can be split into two doubly-excited parts, as can the associated orbital energy denominator, $D_4 = \epsilon_{i} +\epsilon_{j} +\epsilon_{k} +\epsilon_{l}-\epsilon_{a} -\epsilon_{b}-\epsilon_{c} -\epsilon_{d} = (\epsilon_{i} +\epsilon_{j} -\epsilon_{a} -\epsilon_{b})+(\epsilon_{k} +\epsilon_{l}-\epsilon_{c} -\epsilon_{d}) = D_2^A + D_2^B$. The index groups $A$ and $B$ are associated with the two $W_N$ fragments. Due to the symmetry of $T_4$, however, we can associate the index groups to the fragments in either order. Finally, we introduce the identity as $D_4^{-1} D_4$. This leads directly to the factorization
\begin{align}
    \braket{0|T_2^{(1)\dagger} W_N T_4^{[3]}|0}
    &=\frac{1}{2}\langle 0|\left( W_N^A (D_2^A)^{-1} W_N^B D_4^{-1} D_4 T_4^{[3]} \right. + \\
    &\hphantom{=\frac{1}{2}\langle 0|\bigg(} \left. W_N^B (D_2^B)^{-1} W_N^A D_4^{-1} D_4 T_4^{[3]}\right)_C|0\rangle \\
    &= \frac{1}{2}\braket{0|\left( W_N^A (D_2^A)^{-1} W_N^B (D_2^B)^{-1} D_4 T_4^{[3]} \right)_C|0} \\
    &= \frac{1}{2}\braket{0|T_2^{(1)\dagger}T_2^{(1)\dagger} \left(W_N\frac{T_2^2}{2}+W_NT_3 \right)_C|0}
\end{align} where the factor of 1/2 arises because of the two equivalent orderings, and we have used the fact that $D_4=D_2^A +D_2^B$. From \eqref{eq:T4eqn}, the application of $D_4$ to $T_4^{(3)}$ ``undoes'' the four-electron resolvent and removes the denominator coupling of the eight external indices. Kucharski and Bartlett used this transformation as a basis to propose what they referred to as the (\Qf{}) correction,\cite{Kucharski-AQC-1986,kucharski1998efficient} which is given to be
\begin{align}
    E_{(Q_f)}^{[5]}= \frac{1}{2}\braket{0|T_2^{\dagger}T_2^{(1)\dagger}\left(W_N\frac{T_2^2}{2}+W_NT_3 \right)_C|0}.
\end{align}
where the reintroduction of the full $T_2^\dagger$ provides additional higher-order terms. It should be noted that despite the notation (\Qf{}) is a factorized version of the [Q] (``bracket Q'') correction in \eqref{eq:sqrBrakQ}, rather than the (Q) (``parentheses Q'') correction described below. An overview of the factorization theorem as it pertains to coupled-cluster theory is discussed elsewhere.\cite{bartlett2024perspective}

\begin{figure}%{0.6\textwidth}
\includegraphics[scale=0.25]{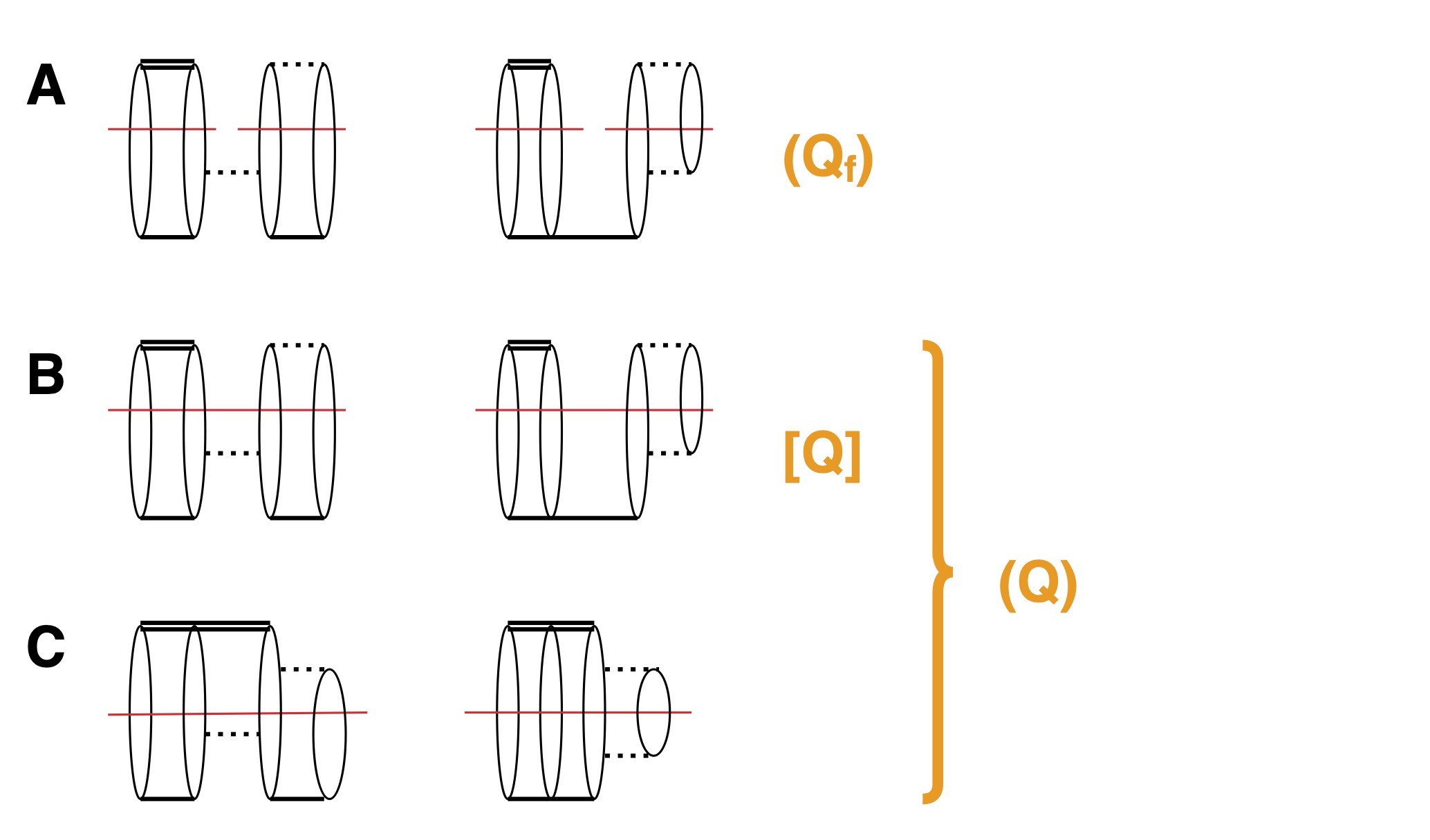}
\caption{Comparison of the pertinent diagrams defining various, perturbative approximations targeting quadruple excitations. Red, horizontal lines indicate denominators. A) The two diagrams defining the (\Qf{}) approximation. B) The two diagrams defining the [Q] approximation. C) These two diagrams in addition to those in B) define the (Q) approximation. In the (Q)${}_\Lambda$ approximation the top $T^\dagger$ vertices in (Q) are replaced by $\Lambda$. }
\label{fig:Qmethods}
\end{figure}

What we refer to as the (Q) correction includes an additional term to those of \eqref{eq:sqrBrakQ}. This is shown to be\cite{kucharski1993coupled,kucharski1998sixth,Bomble-JCP-2005}
\begin{align}\label{eq:ParenthQ}
    \Delta E_{(Q)} = \Delta E_{[Q]} + \braket{0|T_3^{\dagger}W_N T_4^{[3]}|0} 
\end{align}
with $T_4^{[3]}$ given by \eqref{eq:T4eqn}. The distinction between these methods is shown diagramatically in Fig.~\ref{fig:Qmethods}. Note that this extra term would show up as a sixth-order correction to the energy if taking an expectation value coupled cluster (XCC) viewpoint.\cite{bartlett1988expectation,noga1989coupled,bartlett1989alternative} Additionally, this term is not factorizable---meaning there is no way to improve the $\mathcal{O}(N^9)$ scaling of the second term in \eqref{eq:ParenthQ} which is shown diagramatically in Fig.~\ref{fig:Qmethods}C. A final non-iterative quadruples correction is given by\cite{Bomble-JCP-2005}
\begin{align}\label{eq:ParenthQLambda}
    \Delta E_{(Q)_\Lambda} = \braket{0|\Lambda_2 W_NT_4^{[3]}|0} + \braket{0|\Lambda_3 W_N T_4^{[3]}|0} 
\end{align}
where $\Lambda$ are the left-hand wavefunction components of the coupled cluster ground state.

The complete coupled cluster methods and total correlation energies defined by these various approximation, as used in this work, are CCSD(TQ) ($E_\text{CCSD(TQ)} = E_\text{CCSD(T)} + \Delta E_\text{[Q]}$), CCSD(TQ${}_\text{f}$) ($E_\text{CCSD(T\Qf{})} = E_\text{CCSD(T)} + \Delta E_\text{(\Qf{})}$), CCSDT[Q] ($E_\text{CCSDT[Q]} = E_\text{CCSDT} + \Delta E_\text{[Q]}$), CCSDT(Q) ($E_\text{CCSDT(Q)} = E_\text{CCSDT} + \Delta E_\text{(Q)}$), and CCSDT(Q)${}_\Lambda$ ($E_\text{CCSDT(Q)${}_\Lambda$} = E_\text{CCSDT} + \Delta E_\text{(Q)${}_\Lambda$}$).

\subsection{Higher-Level Correlation Predictors}
This work attempts to provide a quantitative means by which to estimate the correlation energy missing at the conclusion of a CCSD(T) calculation. As recent work has demonstrated the capabilities of CCSDT\pQL{} to act as a second Pauling point in the CC expansion (the ``platinum standard'' in comparison to CCSD(T)'s ``gold standard''), we will define the residual higher-level correlation (HLC) as $\Delta E_\text{HLC} = E_\text{CCSDT\pQL{}} - E_\text{CCSD(T)}$. Previous diagnostics of the multi-reference nature of a given molecule are not generally correlated with a quantifiable prediction of the influence of HLC effects to, say, a bond energy. An exception to this, however, is the work of Goodson in 2002\cite{Goodson-JCP-2002}, which was later extended by Schr\"{o}der and coworkers in 2015\cite{Schroder-ZfPC-2015} and made use of in some FPD model chemistries\cite{Feller_JCP_2003, Feller-JCP-2007, Feller-JCP-2017}.

Goodson used continued fractions and truncated Pad\'{e} approximants to derive three formulas for predicting the post-CCSD(T) electron correlation energy $\Delta E_\text{HLC}$:
\begin{align}
    \label{eq:goodson_cf}
    \Delta E_{cf} &= \frac{E_\text{CCSD}^2}{E_{0}} + 
    \frac{\Delta E_\text{(T)}^2}{E_\text{CCSD}} +
    2\frac{E_\text{CCSD}\Delta E_\text{(T)}}{E_0} \\
    \label{eq:goodson_r}
    \Delta E_{R} &= \frac{\Delta E_\text{(T)}^2}{E_\text{CCSD}} + 
    \frac{\Delta E_\text{(T)}^3}{E_\text{CCSD}^2} \\
    \label{eq:goodson_q}
    \Delta E_{Q} &= 2\frac{\Delta E_\text{(T)}^2}{E_\text{CCSD}} + 
    5\frac{\Delta E_\text{(T)}^3}{E_\text{CCSD}^2} 
\end{align}
where $E_0$ is the SCF reference energy, $E_\text{CCSD}$ is the CCSD correlation energy, and $\Delta E_\text{(T)}$ is the perturbative triples contribution from CCSD(T). We will refer to these methods, in order, as CF, R, and Q. In this work, we will also apply these approximations to CCSD(T\Qf{}), in which case we define $\Delta E_\text{(T\Qf{})} \equiv E_{\text{CCSD(T\Qf{})}} - E_{\text{CCSD}} = \Delta E_\text{(T)} + \Delta E_\text{(\Qf{})}$ and use that in place of $\Delta E_\text{(T)}$.

It should be noted that Goodson's continued fraction formula \eqref{eq:goodson_cf} could be extended to take the (\Qf{}) correction as a separate contribution in a similar manner to  Schr\"{o}der et al.\cite{Schroder-ZfPC-2015}, although we have not done so here.  

A better-known alternative to Goodson's continued fraction approach is the \%TAE[(T)] metric of Martin. \cite{Karton-JCP-2006, Martin-MP-2014} Specifically devised in the context of total atomization energies, which, at the time, were the primary means of determining ab initio predictions of enthalpies of formation of small molecules, the \%TAE[(T)] is determined by calculating the contribution of the perturbative triples portion of CCSD(T) to an atomization energy of some species $X$,
\begin{align}\label{eq:percent_t_TAE}
    \%TAE[(T)](X) \equiv \frac{N_C \Delta E_\text{(T)}(C) + N_N \Delta E_\text{(T)}(N) + N_O \Delta E_\text{(T)}(O) + \dots - \Delta E_\text{(T)}(X)}{N_H E_{SCF}(H) + N_C E_{CC}(C) + N_N E_{CC}(N) + \dots - E_{CC}(X)},
\end{align}
where $N_H, N_C, N_N, N_O$ are the number of hydrogen, carbon, nitrogen, and oxygen atoms in the species $X$, $\Delta E_\text{(T)}(X)$ is perturbative triples contribution to the energy of a given species (as above), and $E_{CC}(X)$ is the full CCSD(T) energy (both SCF and correlation) of a given species. This method has been demonstrated to be a good predictor of when effects post-CCSD(T) are important in model chemistries based upon total atomization energies\cite{Karton-JCP-2006}, but is not usually extended to general chemical reactions, a point that will be addressed later in this manuscript.

\section{Methods}

\subsection{Chemical Test-Suite}
In order to benchmark the errors associated with the factorization of the $D_4$ denominator via CCSD(T\Qf{}), we have constructed two test-suites drawn from the singlet species of the W4.17 dataset\cite{Karton-JCC-2017} containing the atoms H, B, C, N, O, and F. These species span a range of chemical properties and bonding---from those well described by single reference methods, such as water, to those of decidedly multi-reference character, such as \ce{BN} and \ce{C2}---and total 93 individual molecules. The first test-suite is composed of the absolute correlation energies of these compounds. This metric contains information about the trueness\cite{Ruscic-IJQC-2014} of the quantum chemical methods applied to correlation energies without fortuitous or designed cancellation of errors, but is applied to properties of molecules that cannot be determined experimentally.

The second test-suite, the ``reaction'' suite, is constructed by extending the ANL-$n$ scheme for determining enthalpies of formations of molecules.\cite{Klippenstein-JPCA-2017} In that work, the enthalpy of formation of a molecule containing only H, C, N, and O atoms was obtained by balancing a chemical reaction where the reactants (LHS) contains the species of interest, the products (RHS) contains an appropriate number of \ce{CH4}, \ce{NH3}, and \ce{H2O} as ``reference species'' to balance the atom count of the LHS, and additional \ce{H2} molecules are added to the reactants or products in order to balance the remaining H atoms. This work extends this scheme in two ways. First, we add \ce{B2H6} and \ce{HF} as reference species for the B and F atoms. Second, we extend the scheme to include non-reference molecules in not just the reactants, but also in the products of the chemical reaction, which are then balanced with the species of the reference set (\ce{H2}, \ce{B2H6}, \ce{CH4}, \ce{NH3}, \ce{H2O}, and \ce{HF}). By considering all unique pairs of non-reference molecules as reactants/products, this test-suite is expanded from a comparatively small number of species to a much more significant number of chemical reactions (93 species and 2859 reactions in the case of the cc-pVDZ basis set,\cite{Dunning-JCP-1989} and 73 species and 1600 reactions in the case of the cc-pVTZ basis set\cite{Dunning-JCP-1989}). The reactions contained within span a considerable range of chemical space, from simple decompositions such as \ce{H2CCH2 \rightarrow HCCH + H2 } to complicated reactions such as \ce{BF3 + NH3 \rightarrow BN + 3HF}, which preserve little-to-no bonding from the products to reactants. The species, reactions, and raw energetic data for all results presented below are may be generated from the data and scripts in the Electronic Supplementary Information.

It should be noted that great care must be applied to the training or fitting of any model to the information contained within this reaction energy test-suite. The reaction energies contained within will be significantly, though subtly, influenced by the set of reference species chosen to balance the left and the right hand side of the chemical reactions. That is, a great majority of the reactions in this reaction energies suite will be connected to a subset of reference molecules containing only six species, and any model trained to minimize errors in the chemical behavior may fall dangerously close to overfitting any information that correlates with features of these six molecules. In the language of machine learning, we would term the absolute energy test-suite as the ``training'' set, and the reaction test-suite as the ``validation'' set.\cite{goodfellow2016deep} It is perhaps informative that the former in our case is far smaller than the latter; indicative of our desire to remain as ab initio as possible in our approach and to avoid (or at least minimize) the aforementioned dangers of fitting the specific reaction scheme utilized.

\subsection{Software}
The closed-shell data generated in this work was obtained via the {\sc NCC} module of the {\sc CFOUR} quantum chemistry package,\cite{Matthews-JCP-2020} using the current development version. All calculations were sufficiently converged such that numerical noise is several orders of magnitude below the statistics presented here. The (\Qf{}) contributions for the open-shell atoms were determined using a Python package designed for rapid prototyping of coupled cluster methods,\cite{UT2} which in turn relies on an interface to {\sc PySCF} for one and two-body integrals.\cite{sun2018pyscf} The open-shell CCSDT\pQL{} calculations were performed using the general CC code, {\sc MRCC}, of K\'{a}llay.\cite{Kallay-JCP-2020} All post-SCF calculations include only valence correlation. 

\subsection{Basis Sets}
The basis sets tested include Dunning's cc-pVDZ and cc-pVTZ,\cite{Dunning-JCP-1989} with the majority of the results obtained using the small cc-pVDZ basis set. This is a result of three independent considerations. First, the effects of quadruple excitations in the majority of high-accuracy model chemistries are determined using this basis set, even if the increment from CCSD(T) to CCSDT is often treated with a cc-pVTZ or larger basis. Second, the purpose of this work is not to replace the HLC terms of these model chemistries, but rather to demonstrate that the magnitude and sign of these corrections can be  estimated using perturbation theory, and cc-pVDZ is sufficient for this purpose. Third, it has been previously reported that CCSDT[Q] displays basis-set dependence discordant with that of CCSDT(Q) and CCSDT\pQL{}, accumulating undesirable errors with basis sets larger than cc-pVDZ.\cite{Bomble-JCP-2005} This may need to be accounted for if one were to attempt to include the fifth-order (\Qf{}) and [Q] corrections in a model chemistry, but as our purpose is simply to provide guidelines for estimating HLC contributions, we will put this issue aside. 

\subsection{Analysis}
The statistical analysis of the absolute and reaction energy test-suites discussed above was performed with the R language,\cite{R-2024} with scripts available in the SI. Each statistical measurement reported below was further characterized by BCa\cite{Efron-JASA-1987, boot-book} analysis of ten-thousand bootstrap replicas\cite{Efron-SS-1986} in order to establish 95\% confidence intervals, all performed using the boot package\cite{R-boot, boot-book} in R version 4.3.3 (2024-02-29). As we are most interested in the ``trueness'' of the methods investigated here\cite{Ruscic-IJQC-2014}, we define the root-mean squared trueness (RMST) as
\begin{equation}\label{eq:rmst}
    \sigma_T \approx \sqrt{\frac{\sum_{i=1}^{N}\left(x_i^2 \right)}{N-1}},
\end{equation}
which is identical to the usual standard deviation formula with the average measurement $\bar{x}$ taken as zero.

\section{Results and Discussion}

%%%%%%%%%%%%%%%%%%%%%%%%%%%%%%%%%%%%%%%%%%%%%%%%%%%%%%%
% FACTORIZATION ERRORS
\subsection{Factorization Error: CCSD(TQ) vs CCSD(T\Qf{})}

\begin{figure}
    \centering
    \includegraphics[width=\textwidth]{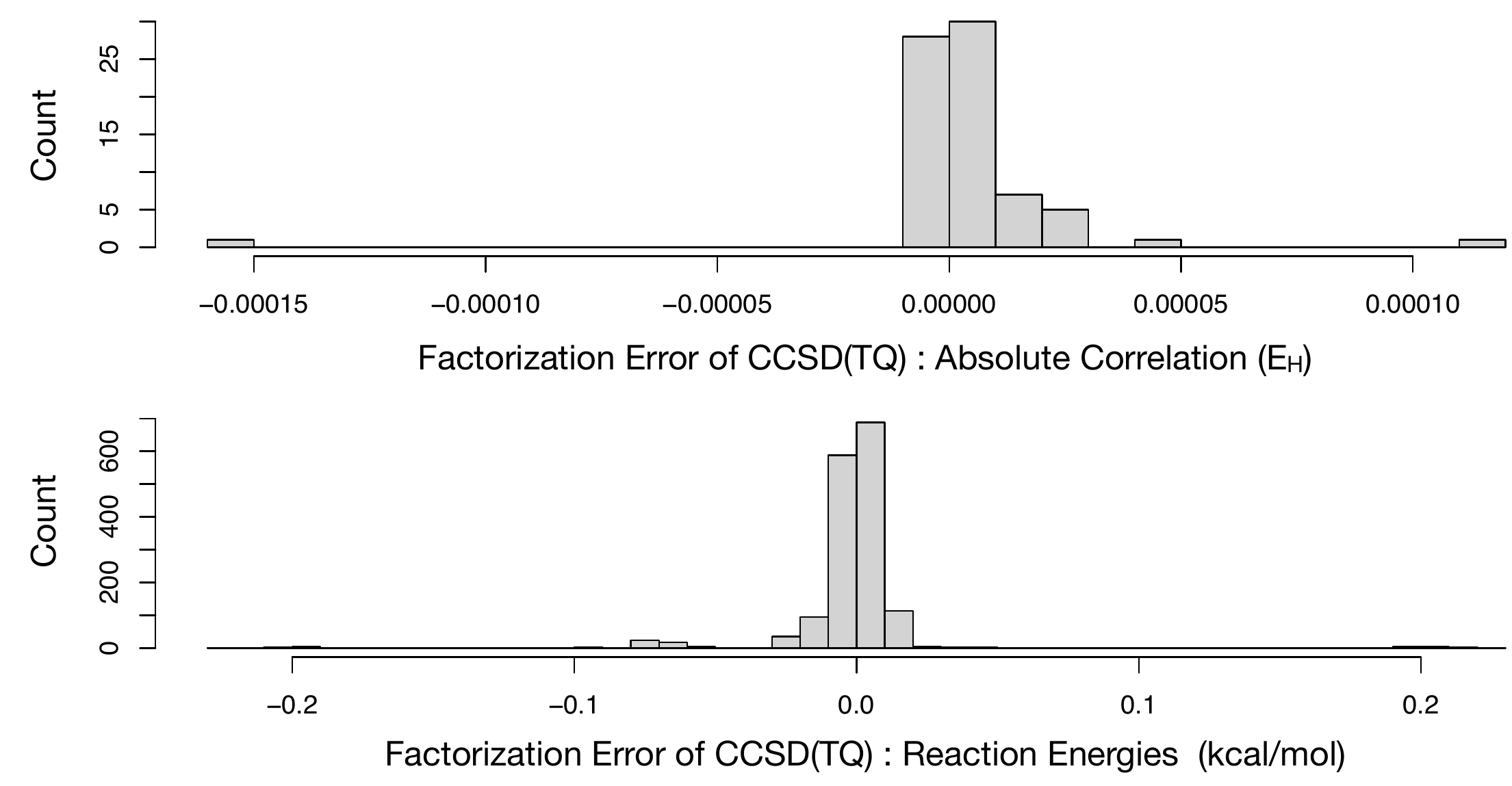}
    \caption{Histogram of the error incurred by factorizing the $D_4$ denominator in CCSD(TQ)/cc-pVDZ correlation energies (top) and contributions to reaction energies (bottom). Outliers on the left and right hand side of the x-axis are BN and C$_2$ (top) and reactions involving these species (bottom).} \label{fig:historgramQfvsQ}
\end{figure}

To characterize errors that might arise by factorizing the $D_4$ denominator as in the CCSD(T\Qf{}) method, our initial results focus on the raw correlation energy differences between the fifth-order [Q] and [\Qf] corrections. The top panel of Fig.~\ref{fig:historgramQfvsQ} displays a histogram of the difference between the fifth-order [Q] and (\Qf{}) correlation energies applied to the absolute energy test-suite calculated with the cc-pVDZ basis set. As shown by the histogram itself, the CCSD(T\Qf{}) and CCSD(TQ) energies coincide to a remarkable degree. This analysis is corroborated in Table~\ref{tab:Qf_vs_Q}, which records an average error of 4.2 $\mu$\EH{}, with an asymmetric 95\% confidence interval that skews toward (\Qf{}) slightly under-predicting [Q]. The reported RMST value of 24.9 $\mu$\EH{} is exceedingly small, and the only outliers are \ce{BN} and \ce{C2}, which are particularly nasty species in this test-set. These results corroborate previous findings in the literature, which demonstrated small errors incurred by this factorization in smaller datasets.\cite{kucharski1998efficient,kucharski1998noniterative,musial2005critical}

This trend continues when analyzing the impact of these quadruples' corrections to reaction energies of the reaction energy test-suite calculated with a the cc-pVDZ basis, as illustrated in the bottom panel of Fig.~\ref{fig:historgramQfvsQ}. Here, we again find a tight error distribution centered around a mean value -1.5 \cal{}, with outliers corresponding to reactions containing \ce{C2} and \ce{BN}. As shown in Table~\ref{tab:Qf_vs_Q}, the 95\% confidence interval is very tight, with an RMST of 28.1 \cal---entirely negligible on the scale of chemical accuracy (1 k\cal). 

Results for the smaller cc-pVTZ test-suites for molecular correlation energies and reaction energies track with those found at the cc-pVDZ level.  We find that (\Qf{})/cc-pVTZ tends to underestimate both raw correlation and reaction energies. Table~\ref{tab:Qf_vs_Q} documents the mean error in the prediction of raw correlation energies to be 7.95 $\mu$\EH{} using the larger cc-pVTZ basis set, with an RMST of 22.9 $\mu$\EH{}. Reaction energies determined using the cc-pVTZ basis set tend to slightly increase the mean error to -2.6 \cal{} and decrease the RMST to 22.4 \cal{}, compared to cc-pVDZ. There is virtually no impact in RMST between the cc-pVDZ and cc-pVTZ basis sets, which seems to indicate that the factorization of [Q] into (\Qf{}) is not a highly basis-set dependent approximation, even if the [Q] correction itself is.

In summary, we find that the errors arising from the factorization of $D_4$ in CCSD(TQ) to be manifestly negligible in both the description of correlation energy and in correlation energy contributions to reaction energies.

\begin{table}[h!]
    \centering
    \setlength{\extrarowheight}{1mm}
    \caption{Mean signed error (MSE), RMST, and max absolute error (MAXE) values incurred by the factorization of the $D_4$ denominator in CCSD(T\Qf{}) for both absolute correlation energies and the test-suite of reaction energies. Subscripts and superscripts correspond to 95\% confidence intervals determined via BCa analysis of 10,000 bootstrap samples.}\label{tab:Qf_vs_Q}\sisetup{table-format = 2.8, table-alignment-mode = format, table-align-text-after = false}
    \begin{tabular}{l S S}
    \hline
    & \multicolumn{2}{c}{Absolute Energy ($\mu$\EH{})} \\
    \hline
    & \multicolumn{1}{c}{cc-pVDZ} & \multicolumn{1}{c}{cc-pVTZ} \\
    MSE  & 4.2${}_{-7.0}^{+4.5}$ & 8.0${}_{-4.5}^{+5.4}$  \\
    RMST        & 24.9${}_{-14.4}^{+19.2}$ & 22.9${}_{-9.1}^{+16.3}$ \\
    MAXE         & 154.2 & 129.9 \\
    \hline
    &  \multicolumn{2}{c}{Reaction Energy (\cal{})} \\
    \hline
    & \multicolumn{1}{c}{cc-pVDZ} & \multicolumn{1}{c}{cc-pVTZ} \\
    MSE  &  -1.5${}_{-1.3}^{+1.4}$   & -2.6${}_{-1.1}^{+1.1}$ \\
    RMST &  28.1${}_{-3.9}^{+4.6}$   & 22.4${}_{-2.5}^{+2.7}$ \\
    MAXE & 227.5                     & 168.9\\
    \hline
    \end{tabular}
\end{table}

%%%%%%%%%%%%%%%%%%%%%%%%%%%%%%%%%%%%%%%%%%%%%%%%%%%%%%%
% HLC PREDICTION
\subsection{Scaled-Perturbation Estimators of Higher-Level Correlation}
\begin{figure}[h!]
    \centering
    \includegraphics[width=\textwidth]{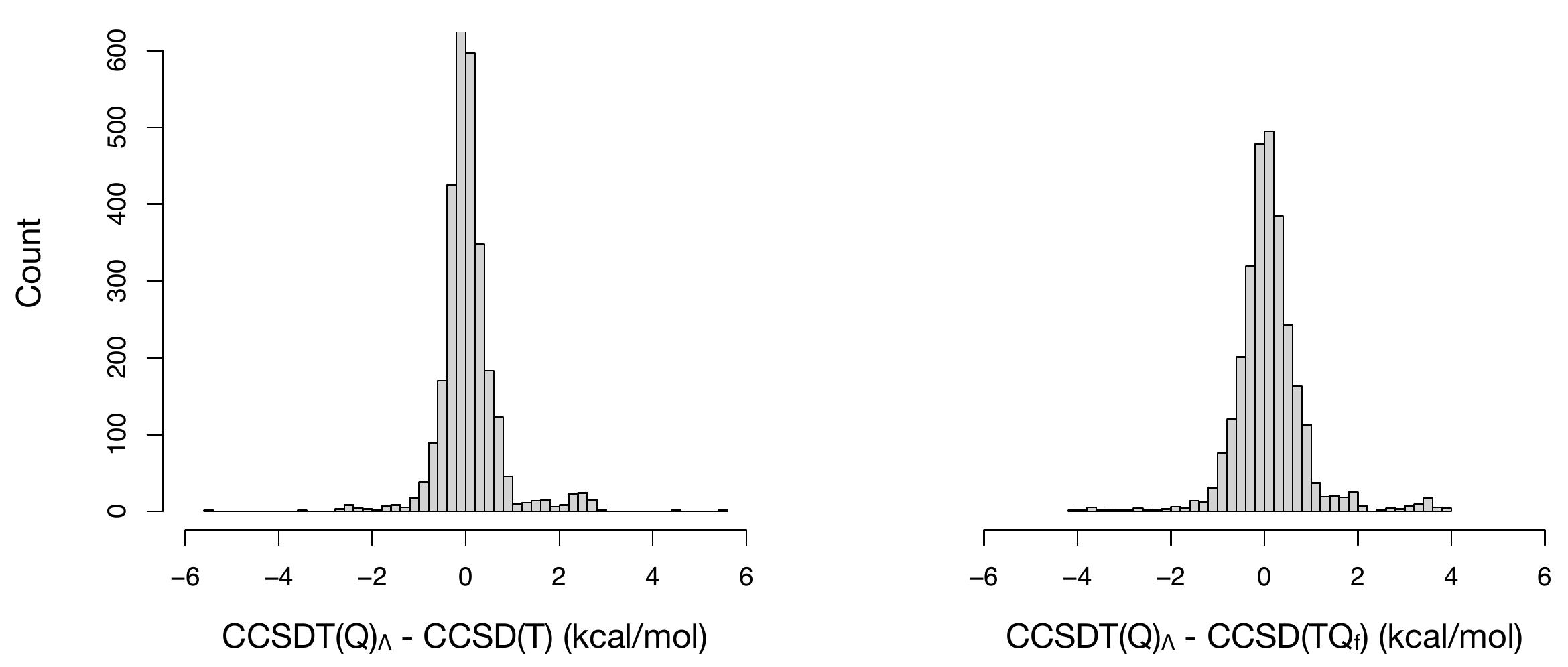}
    \caption{Histogram of differences between CCSDT\pQL{} and CCSD(T)/cc-pVDZ (left), CCSD(T\Qf{})/cc-pVDZ (right) for reaction energies.}\label{fig:HLC_t_vs_tqf_rxn_dz}
\end{figure}

Having determined that the factorization approximations in CCSD(T\Qf{}) introduce no significant errors on the scale of \kcal{} accuracy, we may being to consider what role these methods might play in model chemistries. To this end, Fig.~\ref{fig:HLC_t_vs_tqf_rxn_dz} displays histograms of the difference between CCSDT\pQL{} (our ``platinum standard'' method for post-CCSD(T) calculations) and CCSD(T)/CCSD(T\Qf{}) in the context of cc-pVDZ  contributions to reaction energies. The mean signed error, RMST, and max absolute errors are also tabulated in Table~\ref{tab:tqf_vs_t_dz_rxns} (under the ``None'' column, indicating that no additional correction or extrapolation is applied). In general, we find that CCSD(T) displays smaller mean HLC contributions than CCSD(T\Qf{}) (0.05 vs 0.10 \kcal{}), along with reduced RMST (0.66 vs 0.77 \kcal{}) of the missing HLC, but with a larger maximum errors (5.53 vs 4.00 \kcal{}).  While CCSD(TQ) has been shown to improve the treatment of total correlation energies\cite{kucharski1998efficient,kucharski1998noniterative,musial2005critical} (a result we have also observed in the process of performing this study), the inclusion of fifth-order quadruples corrections at the CCSD level of theory does not improve the prediction of reaction energies, and in fact harms it.  

However, a more prescient application of CCSD(T\Qf{}) is not as a direct substitute for more complete theories such as CCSDT\pQL{}, but rather as a predictor of them. For instance, we may ask the following question: ``given just a CCSD(T)-cost calculation, might it be possible to estimate the size of the missing HLC terms to, say, one or two \kcal{}''? Tacit within our attempt to address this question are three critical approximations: that the cc-pVDZ basis set is sufficient to represent post-CCSD(T) correlation (model chemistries aiming for \kcal{} accuracy may find this  cumbersome and model chemistries aiming for \kJmol{} accuracy may find this insufficient), that CCSDT\pQL{} is a sufficient representation of correlation effects after CCSD(T) (evidence indicates that this is true for well-behaved species, but it may not be sufficient for multireference molecules such as \ce{C2} and \ce{BN}), and that core-valance correlation beyond CCSD(T) may be be safely neglected for accuracy on the scale of one to two \kcal{} (which is certainly true for molecules containing atoms $Z\leq10$, but not so obviously true for heavier atoms). 

\begin{figure}[h!]
    \centering
    \includegraphics[width=\textwidth]{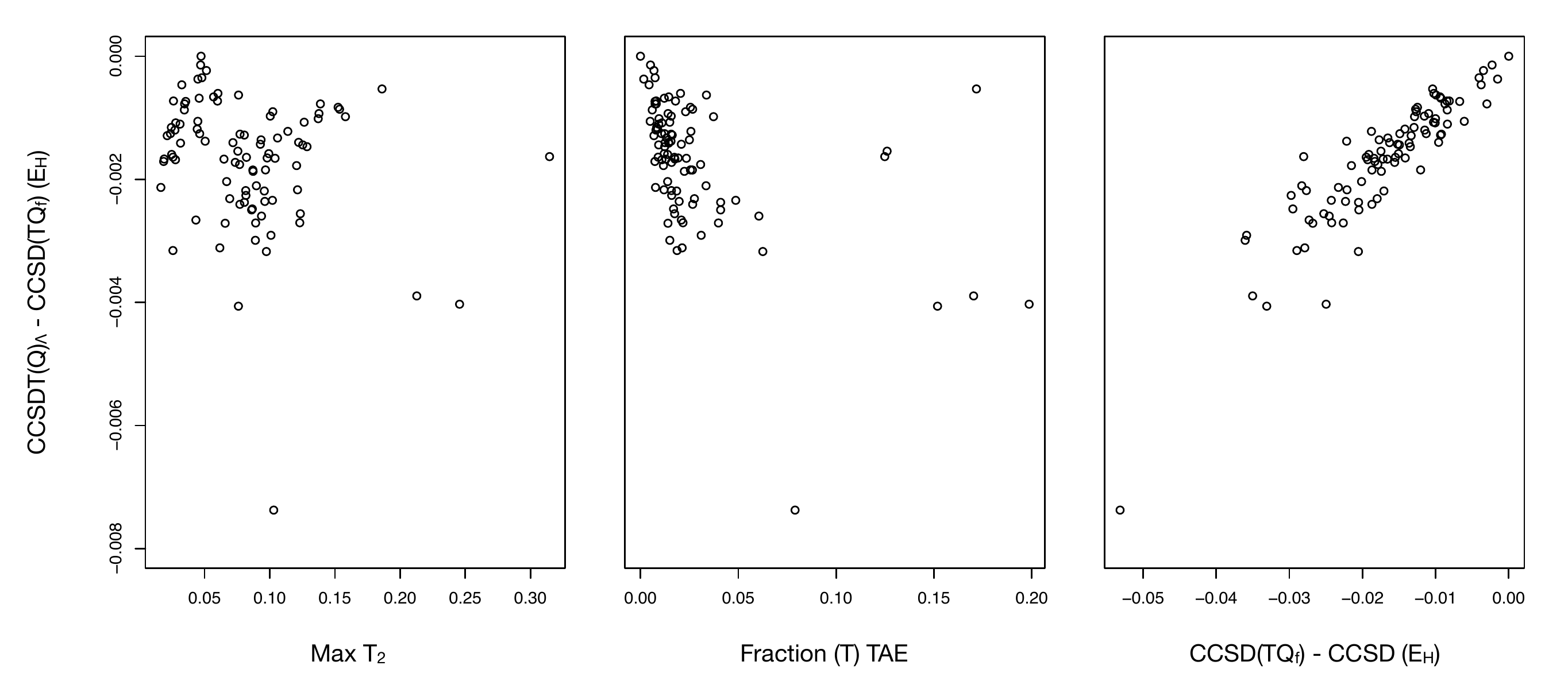}
    \caption{Max $T_2$ (left) and \%TAE[(T)] (middle, fractional form) multireference diagnostics as predictors of higher-level correlation, compared against the (T\Qf{}) perturbative corrections (right) suggested in this work.}\label{fig:MRI_vs_SPE}
\end{figure}

Related versions of this question have been asked before, specifically in the pursuit of so-called multi-reference indices (MRI).\cite{Karton-JCP-2006, Martin-MP-2014, lee1989diagnostic, lee1989theoretical, ivanov2005new, bartlett2020index, faulstich2023s} But, while these indices are useful guides to the degree of complexity in the electronic wavefunction, it is not always obvious how to translate this into quantitative prediction of HLC effects. This is demonstrated in Fig.~\ref{fig:MRI_vs_SPE}, where the often quoted MRI, the max CCSD $T_2$, is plotted on the x-axis and the missing HLC of CCSD(T\Qf{})/cc-pVDZ is plotted on the y-axis. Despite the fact that this index certainly contains some information about the suitability of the CCSD wavefunction to address these molecular correlation energies, this index does not track in a particularly quantitative manner with extent of correlation missing in CCSD(T). Quantitative correlation is certainly improved by using Martin's \%TAE[(T)], displayed in the central panel of Fig.~\ref{fig:MRI_vs_SPE}, which show a generally linear relationship between the fraction of the TAE that is accounted for by the perturbative triples correction and the extent of the the HLC corrections. The agreement is not perfect---the trend appears to break down for particularly large \%TAE[(T)]---and it is not generalized to arbitrary molecular reactions, but it is certainly an improvement. We can, however, go further. The right-most panel of this figure employs a different type of metric that is intimately related to the \%TAE[(T)] of Martin. However, instead of taking a fraction of an atomization energy (a quantity which blends features of the absolute and reaction energy measures), we display the relationship between post-CCSD(T\Qf{}) correlation energy and simply the difference between CCSD(T\Qf{}) and CCSD in a cc-pVDZ basis set. Said otherwise, rather than examining the fraction of an atomization energy that is accounted for by perturbative triples, we examine the total of a molecule's correlation energy that is accounted for by perturbative triples and quadruples. This approach reveals a linear relationship between the HLC correction to CCSD(T\Qf{}) and the (T\Qf{}) correction to CCSD, and may thus be modeled as by a simple equation of the form $y \approx b*x$, where $b$ is the slope obtained by least squares fit to the data in the rightmost panel of Fig.~\ref{fig:MRI_vs_SPE}. 

This has a particularly intuitive interpretation. Let us begin by writing out an expression for the approximation of CCSDT\pQL{} using this linear fit (assuming all terms are evaluated with the same basis set). 
\begin{align}\label{eq:spe}
    E_{\text{CCSDT(Q)}_\Lambda} &\approx E_{\text{CCSD(T\Qf{})}} +  b(E_{\text{CCSD(T\Qf{})}} - E_{\text{CCSD}} ) \\
    &\approx E_{\text{CCSD}} + (1 + b)\Delta E_\text{(T\Qf{})}
\end{align}
The above equation claims that the difference between CCSD and CCSDT\pQL{} may be captured by the leading-order perturbative treatment of ${T}_3$ and ${T}_4$ from CCSD(T\Qf{}) multiplied by a (hopefully) small scaling factor. We thus term this approach as a ``scaled perturbation estimator'', or SPE for short, and will examine its application in the remainder of this manuscript. Importantly, the SPE scaling factor is not fit to data from the reaction energies, but to the  correlation energies of individual molecules.

\begin{figure}[h!]
    \centering
    \includegraphics[width=0.95\textwidth]{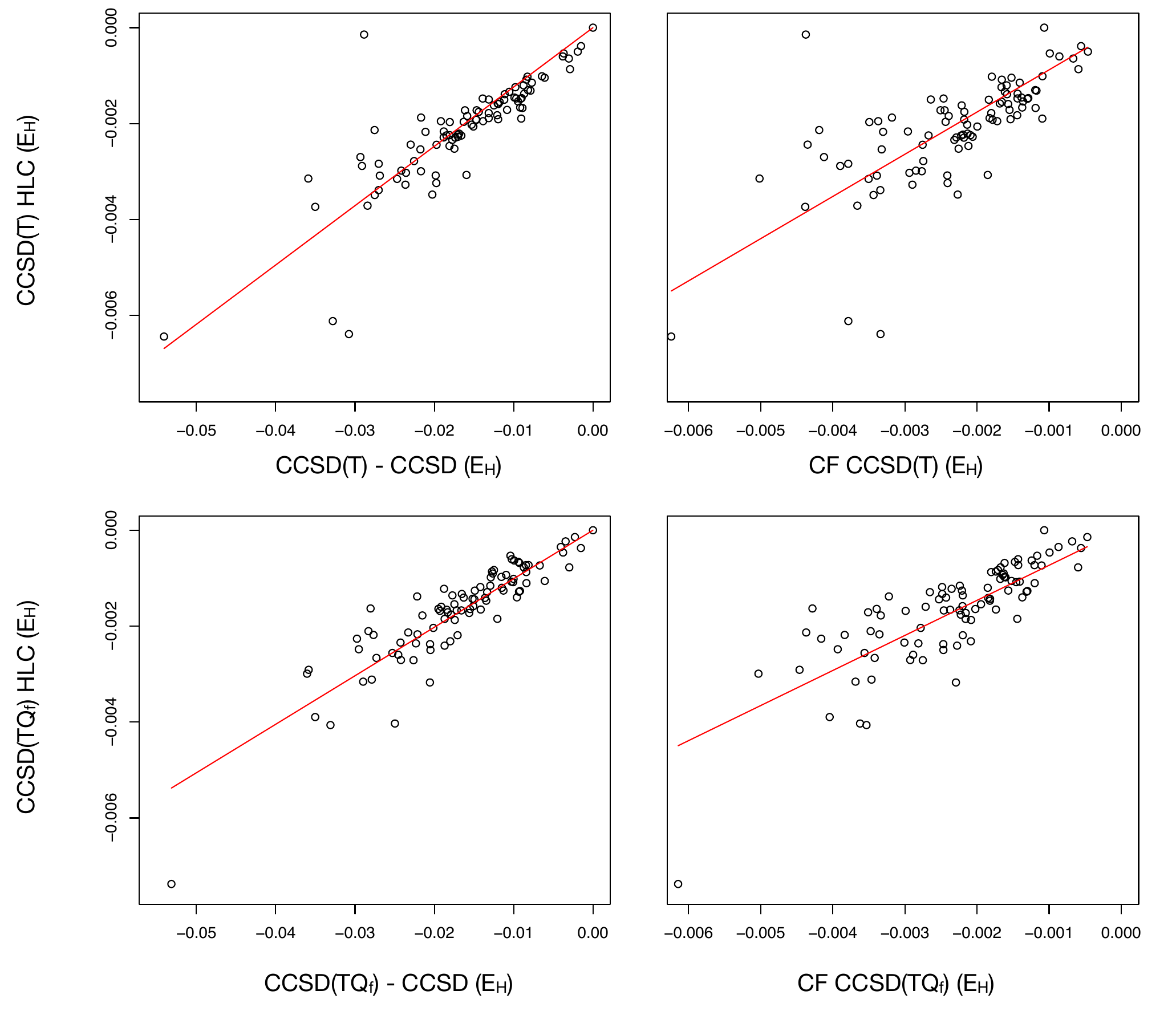}
    \caption{Trends of CCSD(T)-scaling methods vs their HLC corrections in a cc-pVDZ basis set applied to molecular correlation energies. The top row uses CCSD(T) as the base method, and the bottom row uses CCSD(T\Qf{}) as the base method. The red line indicates a linear fit with zero intercept. }\label{fig:goodson_hlc_data_dz}
\end{figure}

Fig.~\ref{fig:goodson_hlc_data_dz} displays two variations of this SPE using either CCSD(T) or CCSD(T\Qf{}) as the ``base'' method. It also displays the relationship of the HLC of CCSD(T) and CCSD(T\Qf{}) with  Goodson's CF approximation, which we find also scales roughly linearly with the size of higher-level correlation contributions to molecular energies. It should be noted that we also tested Goodson's R and Q metrics, but found them inferior to Goodson's CF method. The point of these plots is that the location of each black point on the x-axis can be determined at $\mathcal{O}(N^7)$ cost, while the y-value requires $\mathcal{O}(N^9)$. Thus, a linear fit of the form $y \approx b*x$ to any of these plots could potentially be used as a HLC estimator at a cost no greater than CCSD(T). The coefficients of the linear fits of these trends are displayed in Table~\ref{tab:tqf_vs_t_dz_raw}, and are portrayed by the red lines in Figure~\ref{fig:goodson_hlc_data_dz}. 

\begin{table}[h!]
    \centering
    \setlength{\extrarowheight}{1mm}
    \caption{Parameters of linear fits ($y \approx b*x$) to CCSD(T) and CCSD(T\Qf{}) HLC corrections in a cc-pVDZ basis set, determined from molecular correlation energies using either the ``scaled perturbation estimator'' (SPE) or Goodson's continued fraction (CF) method.\cite{Goodson-JCP-2002}.}\label{tab:tqf_vs_t_dz_raw}\sisetup{table-format = 2.8, table-alignment-mode = format, table-align-text-after = false}
    \begin{tabular}{l l S S}
    \hline
    \multirow{2}{*}{Base Method}& & \multicolumn{1}{c}{SPE} &  \multicolumn{1}{l}{Goodson CF} \\
    & & b & b \\
    \hline
    \multirow{2}{*}{CCSD(T)}        & Best Fit     &0.1235 & 0.8786 \\
    & Std. Err. &  0.0035 & 0.0336 \\
    \multirow{2}{*}{CCSD(T\Qf{})} & Best Fit     & 0.1012 &  0.7309 \\
    & Std. Err. & 0.0025  & 0.0247  \\
    \hline
    \end{tabular}
\end{table}

\begin{figure}[h!]
    \centering
    \includegraphics[width=\textwidth]{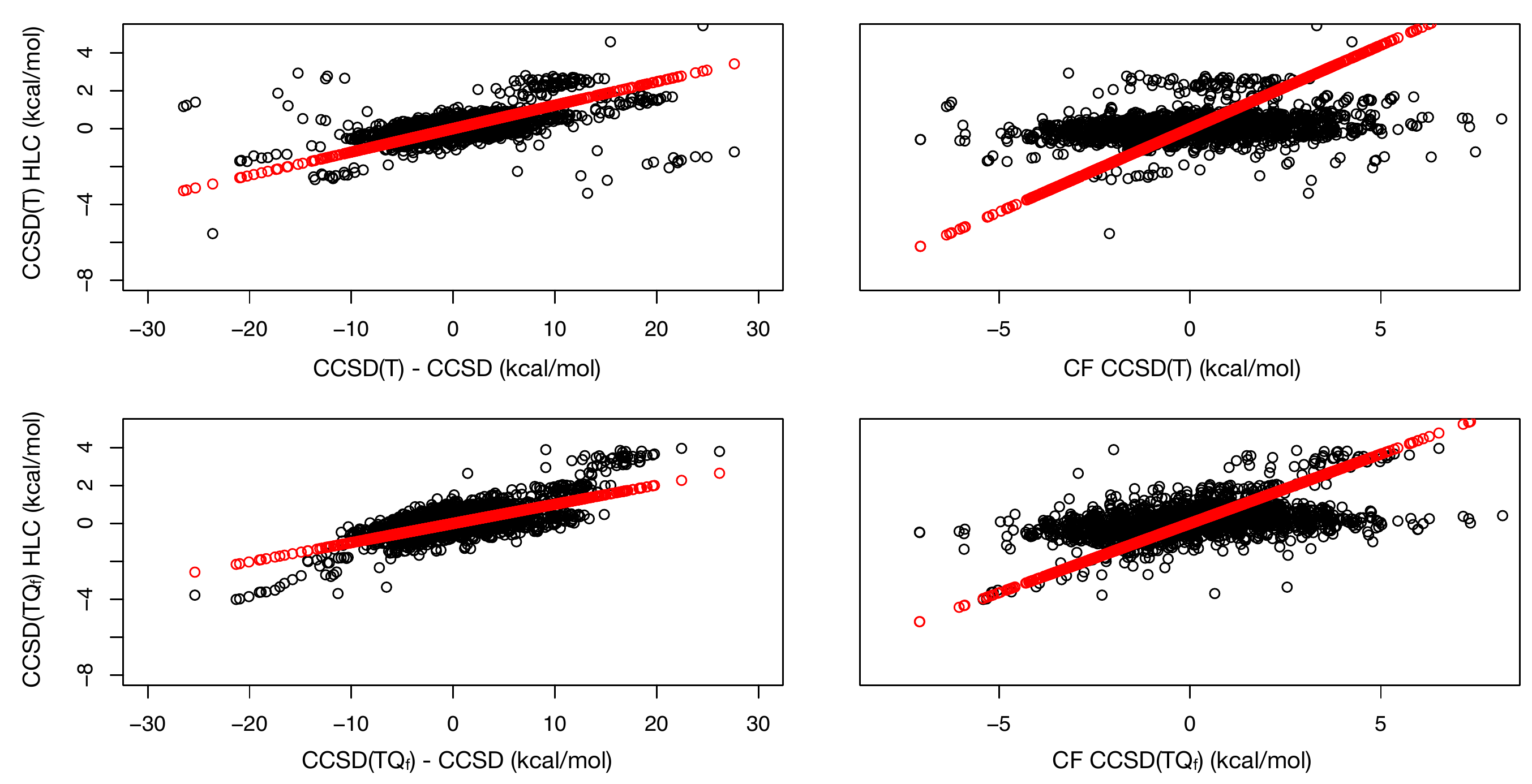}
    \caption{Relationship of CCSD(T) and CCSD(T\Qf{}) vs their HLC corrections with a cc-pVDZ basis set applied to reaction energies. The top row uses CCSD(T) as the base method, and the bottom row uses CCSD(T\Qf{}) as the base method. Red dots indicate the values obtained using the linear fits derived from molecular correlation energies, obtained at CCSD(T) cost.}\label{fig:goodson_hlc_rxns_dz}
\end{figure}

We then examine the performance of six of these models in the prediction of cc-pVDZ HLC in the reaction energies test-suite:
\begin{itemize}
    \item{HLC of CCSD(T) predicted by linear fit of (T),} 
    \item{HLC of CCSD(T) predicted by Goodson CF of CCSD(T),}
    \item{HLC of CCSD(T) predicted by linear fit of Goodson CF of CCSD(T),} 
    \item{HLC of CCSD(T\Qf{}) predicted by linear fit of (T\Qf{}),} 
    \item{HLC of CCSD(T\Qf{}) predicted by Goodson CF of CCSD(T\Qf{}),}
    \item{HLC of CCSD(T\Qf{}) predicted by linear fit of Goodson CF of CCSD(T\Qf{}).} 
\end{itemize}
These results are summarized in Fig.~\ref{fig:goodson_hlc_rxns_dz} and Table~\ref{tab:tqf_vs_t_dz_rxns}, the latter of which also includes the HLC errors accrued in model chemistries that ignore higher-level correlation altogether in the ``None'' column. It is immediately obvious that the fairly reasonable fits in molecular correlation energies do not necessarily translate to even qualitatively accurate predictions of HLC contributions to reaction energies. Take for example, Goodson's continued fraction method aplied to CCSD(T) (top right panel of Fig.~\ref{fig:goodson_hlc_rxns_dz}. Not only does the method itself show no trend with the actual size of the HLC contributions to reaction energies (reflected in a large RMST of 1.79 \kcal{} and max absolute error of 8.70 \kcal{}), the linear fit of this metric to the molecular correlation energies doesn't even qualitatively track the true data in the reaction energies. This is an unfortunately common theme in computational thermochemistry : approximate methods that are reasonable for raw correlation energies are not guaranteed to be reasonable for reaction energies. The SPE approaches based off CCSD(T) and CCSD(T\Qf{}) perform markedly better, with RMST of 0.61 and 0.52 \kcal{}, respectively. We also observe some trends that indicate that correct physics is being accounted for in these approaches: the sign of the perturbative correction to CCSD generally corresponds to the sign of the missing HLC (hence the diagonal slant to this data). Perhaps most surprising is that CCSD(T\Qf{}), which is a less accurate treatment of correlation in the determination of reaction energies than CCSD(T), performs significantly better than any of the other methods considered here when used to inform a simple linear regression. This is particularly reflected in the maximum error exhibited by the CCSD(T\Qf{})-based SPE, which is nearly 2 \kcal{} smaller than the nearest competitor. Visually, we also see that the inclusion of these fifth-order quadruples corrections in CCSD(T\Qf{}) tightens (makes more diagonal) the relationship between the perturbative treatments of $T_3$ and $T_4$ and the higher-level correlation correction, even if the spread of said HLC contributions is larger. That the factorized, fifth-order quadruple corrections in CCSD(T\Qf{}) play such an important role in these predictions has, to the best of our knowledge, not been observed in the extant literature. It also suggests that the claims of equation \eqref{eq:spe} are not unfounded, and that the scaled perturbation of CCSD by (T\Qf{}) may be made into a reasonable estimate for the more rigorous treatment of $T_3$ and $T_4$ in CCSDT\pQL{}. 

That this simple, empirical scaling works so well is a bit surprising, but nonetheless provides a means by which to reasonably estimate the size and sign of HLC contributions in chemical reactions with nothing more than values that may be routinely calculated at CCSD(T)-cost and a little bit of mental math. The answer to the above question: ``Is it possible to estimate higher-level correlation effects to one or two \kcal{} using just CCSD(T)-cost methods?'' seems to have been answered in the affirmative (even accounting for the factor of two increase in the RMST that would better align these values with a 95\% confidence interval), at least for the reactions within the current test-suite. 

It should be reiterated that these models are not necessarily intended to be used as replacements for higher-level correlation in thermochemical recipes, especially not in anything that aims for accuracy on the order of a \kcal{}. Such an application would need to be carefully studied in the context of the individual model chemistry under examination, and would require further developments to reduce the RMST of the SPE predictions of higher level correlation. While the (T\Qf{})-based SPE developed in this work is the best predictor of HLC that we know of, a prediction of a HLC contribution to a reaction energy on the order of 2 \kcal{} should probably be taken as no more than an indication that a more rigorous treatment of post-CCSD(T) correlation is in order if \kcal{} accuracy is desired.  

\begin{table}[h!]
    \centering
    \setlength{\extrarowheight}{2mm}
    \caption{Mean signed error (MSE), RMST, and max absolute errors (MAXE) of various CCSD(T)-cost predictors of higher-level correlation contributions to reaction energies, as described in the main text.  Values in subscripts and superscripts correspond to 95\% confidence intervals determined via BCa analysis of the reported predictor applied to 10,000 bootstrap samples of the reaction test-suite, respectively. All values are in \kcal{}}.\label{tab:tqf_vs_t_dz_rxns}\sisetup{table-format = 2.8, table-alignment-mode = format, table-align-text-after = false}
    \begin{tabular}{c m{2cm} S S S S }
    \hline
    Base Method & & \multicolumn{1}{c}{None} & \multicolumn{1}{c}{SPE} & \multicolumn{1}{c}{Linear CF} & \multicolumn{1}{c}{Goodson CF} \\
    \hline
    \multirow{3}{*}{CCSD(T)} 
    & MSE  & 0.05${}_{-0.02}^{+0.02}$ &  0.06${}_{-0.02}^{+0.02}$ & 0.02${}_{-0.06}^{+0.06}$ & 0.02${}_{-0.07}^{+0.06}$\\
    & RMST & 0.66${}_{-0.04}^{+0.04}$ & 0.61${}_{-0.05}^{+0.06}$ & 1.58${}_{-0.05}^{+0.06}$ & 1.79${}_{-0.06}^{+0.06}$  \\
    & MAXE & 5.53 & 5.04 & 7.80 & 8.70 \\
    \hline
    \multirow{3}{*}{CCSD(T\Qf{})} 
    & MSE  & 0.10${}_{-0.03}^{+0.03}$ & 0.07${}_{-0.02}^{+0.02}$ & 0.08${}_{-0.05}^{+0.04}$ & 0.06${}_{-0.06}^{+0.06}$\\
    & RMST & 0.77${}_{-0.04}^{+0.04}$ & 0.52${}_{-0.02}^{+0.02}$ & 1.21${}_{-0.04}^{+0.04}$ & 1.64${}_{-0.05}^{+0.05}$\\
    & MAXE & 4.00 & 2.97 & 5.57 & 7.78 \\
    \hline
    \end{tabular}
\end{table}

\subsection{Connection to Martin's \%TAE}
Finally, it is worth commenting on the connection between these scaled perturbation estimators and the percent-(T) TAE based methods that are commonly employed. In Karton et al., 2006\cite{Karton-JCP-2006}, it was found that the percentage of connected ${T}_4$ and ${T}_5$ contribution to total atomization energies post-CCSD(T) was reasonably correlated with contribution from (T): $\%TAE[{T}_4 + {T}_5] \approx 0.126 \times \%TAE[(T)]$.  Note that the factor of 0.126 has been obtained by fitting to a set of atomization energies, but is relatively similar to the 0.123 we obtain for the fit of the SPE fit of HLC based on CCSD(T).

Now, this may be simplified to better compare to the methods presented in this work. If we insert the definition of \%TAE[$X$] \eqref{eq:percent_t_TAE}, the denominators may be cancelled such that
\begin{align}
    &N_C \Delta E_\text{HLC}(C) + N_N \Delta E_\text{HLC}(N) + \dots - \Delta E_\text{HLC}(X) \nonumber{}\\
    &\quad\quad \approx 0.126 \times \left(N_C \Delta E_\text{(T)}(C) + N_N \Delta E_\text{(T)}(N) + \dots - \Delta E_\text{(T)}(X)\right),
\end{align}
where $\Delta E_\text{HLC}$ indicates the high-level correlation correction from $T_4$ and $T_5$. Then we may determine an estimate of the HLC contributions for a given reaction, for example, the isomerization of HCN to HNC,
\begin{align}
    \Delta E^{rxn}_\text{HLC}(\ce{HCN} \rightarrow \ce{HNC}) &= \Delta E^{TAE}_\text{HLC}(\ce{HCN)} - \Delta E^{TAE}_\text{HLC}(\ce{HNC)} \nonumber \\
    &\approx 0.126\left( E_{atoms} - \Delta E_\text{(T)}(\ce{HCN})\right) \nonumber \\
    &- 0.126\left(E_{atoms} 
    - \Delta E_\text{(T)}(\ce{HNC}) \right) \nonumber \\
    &= 0.126\left( \Delta E_\text{(T)}(\ce{HNC}) - \Delta E_\text{(T)}(\ce{HCN}) \right), 
\end{align}
where $E^{rxn}_\text{HLC}(\ce{HCN} \rightarrow \ce{HNC})$ is the contributions of HLC to the reaction energy and $E_{atoms}$ is a shorthand for sum of the $\Delta E_\text{(T)}$ for the H, C, and N atoms. Thus, the \%TAE[(T)] metric may actually be used in a nearly identical manner to the scaled perturbation estimators developed in this work, though the fit of the scaling parameters in that context are determined from percentages of atomization energies rather than absolute correlation energies, which can (and does) have significant consequences for the accuracy of the resulting model. The main contributions of this work, then, are the theoretical justification of performance of these scaled perturbation estimators and/or \%TAE[(T)] metrics, the determination and benchmarking of these scaling factors for a set of 93 molecules and 2859 reactions containing atoms H, B, C, N, O and F in a cc-pVDZ basis set, and the demonstration that the factorization of the CCSD(T\Qf{}) method both incurs negligible error for the vast majority of species/reactions contained in these test-suites and noticeably improves the performance of SPE methods for estimating higher-level correlation.

\section{Conclusions}
The development of new tools and methodologies which  circumnavigate the sometimes unaffordable computational cost associated with high-level coupled cluster theory is particularly important in the domain of thermochemistry targeting chemical accuracy. However, there are currently only limited means in which to evaluate the importance of post-CCSD(T) corrections to correlation energies of molecules or reactions \textit{a priori}. Furthermore, the viability of cheaper quadruples corrections like (\Qf{}) have not been holistically assessed in the domain of thermochemistry. The current work takes steps to address these two gaps in the literature. We find that factorizing the fifth-order [Q] diagrams of CCSD(TQ), leading to the (\Qf{}) correction and CCSD(T\Qf{}) method, introduces minimal error and a $\mathcal{O}(N^2)$ reduction of cost in applications to thermochemistry, even when applied to significantly multireference species such as \ce{C2} and \ce{BN}. 

We then examined the use of these methods in the context of higher-level correlation corrections in chemical reaction thermodynamics. While we find that CCSD(T\Qf{}) on its own is not sufficient to account for HLC effects in reaction energies, we demonstrate that these factorized quadruples methods can play an important role in the estimation of post-CCSD(T) correlation energies such as  CCSDT\pQL{}. In particular, we investigate two potential HLC estimators;  Goodson-style continued faction methods and a method we refer to as scaled perturbation estimators. While we did not find the first to be a reasonable route to estimating HLC effects, we find that a simple linear regression of the leading-order perturbative corrections to CCSD in CCSD(T\Qf{}) reliably predicts the size of post-CCSD(T\Qf{}) correlation in a cc-pVDZ basis set, with a mean error of 0.07 \kcal{}, an RMST of 0.52 \kcal{}, and a maximum error of 2.97 \kcal{}. These results improves upon the CF method applied to this same dataset by a factor of more than three. Importantly, we also find that the maximum absolute errors of these predictors are dramatically reduced by the inclusion of the factorized quadruples terms. We conclude by showing the connection between these methods and the \%TAE[(T)] that is commonly used as a multireference index. Given their low cost, these scaled perturbation estimates may be a viable means by which to ballpark the importance of post-CCSD(T) contributions to reactions energies in other studies without expending more than CCSD(T)/cc-pVDZ effort. This provides a useful guide for electronic-structure practitioners who are uncertain if these higher-level effects are of particular relevance in a given reaction.

\begin{acknowledgement}
 D.A.M. and J.H.T. acknowledge support by the US National Science Foundation under grant CHE-2143725. J.H.T. acknowledges support from the SMU Moody School of Graduate and Advanced Studies.  R.J.B and Z.W.W acknowledge support from the Air Force Office of Scientific Research under AFOSR Award No. FA9550-23-1-0118. Z.W.W. thanks the National Science Foundation and the Molecular Sciences Software Institute for financial support under Grant No. CHE-2136142.
\end{acknowledgement}

\begin{suppinfo}
    Additional supplemental information is available as a .zip file, containing all of the
    molecular structures, raw energetic data, and analysis scripts (in R) used to obtain
    the data presented in the manuscript.
\end{suppinfo}

\bibliography{references}

\end{document}